\documentclass[journal]{IEEEtran}

\IEEEoverridecommandlockouts

\usepackage{balance}

\usepackage{physics}
\usepackage{amsfonts}
\DeclareMathOperator*{\argmax}{argmax}

\newcommand{\mi}{\mathrm{i}}

\usepackage{array}
\usepackage{cuted}
\usepackage{flushend}

\usepackage{paralist}
\usepackage{amsmath,accents}
\usepackage{cite}
\usepackage{amsmath,amssymb,amsfonts}
\usepackage{algorithmic}
\usepackage{textcomp}
\usepackage{ stmaryrd }
\usepackage{mathptmx} 
\usepackage[export]{adjustbox}
\usepackage[section]{placeins}
\usepackage{graphicx}
\usepackage{epstopdf}
\usepackage{subfigure}
\usepackage{caption}
\usepackage{times} 
\usepackage{floatrow}
\usepackage{multirow}
\usepackage{multicol}
\usepackage{mathtools}
\usepackage{float}
\usepackage{xcolor}
\usepackage{upgreek}
\usepackage{ mathrsfs }
\usepackage{ bbm }
\usepackage{bm}
\usepackage{cleveref}
\usepackage{lipsum}
\usepackage{widetext}
\usepackage{tabularx, calc}

\DeclareMathAlphabet{\mathcal}{OMS}{cmsy}{b}{n}
\DeclareMathAlphabet{\mathcal}{OMS}{cmsy}{m}{n}

\newtheorem{algorithm}{\indent Algorithm}
\def\BibTeX{{\rm B\kern-.05em{\sc i\kern-.025em b}\kern-.08em
    T\kern-.1667em\lower.7ex\hbox{E}\kern-.125emX}}
\begin{document}
	\title{Simultaneous estimation of parameters and the state of an optical parametric oscillator system*}


\author{Qi Yu$^{1}$, Shota Yokoyama$^{2}$, Daoyi Dong$^{1}$, David McManus$^{2}$ and Hidehiro Yonezawa$^{2}$
\thanks{*This work was supported by the Australian Research Council's Discovery Projects funding scheme under Project DP190101566, the Centre of Excellence CE170100012 and the U.S. Office of Naval Research Global under Grant N62909-19-1-2129.}
\thanks{$^{1}$ Qi Yu and Daoyi Dong are with the School of Engineering and Information Technology,
        The University of New South Wales, Canberra, ACT 2600, Australia
        {\tt\small yuqivicky92@gmail.com;daoyidong@gmail.com.}}%
\thanks{$^{2}$ Shota Yokoyama, David McManus and Hidehiro Yonezawa are with the Centre for Quantum Computation and Communication Technology and School of Engineering and Information Technology,
The University of New South Wales, Canberra, ACT 2600, Australia
        {\tt\small s.yokoyama@adfa.edu.au; d.mcmanus@adfa.edu.au; h.yonezawa@unsw.edu.au.}}%
}
\maketitle

\begin{abstract}
In this paper, we consider the filtering problem of an optical parametric oscillator (OPO). The OPO pump power may fluctuate due to environmental disturbances, resulting in uncertainty in the system modeling. Thus, both the state and the unknown parameter may need to be estimated simultaneously. We formulate this problem using a state-space representation of the OPO dynamics. Under the assumption of Gaussianity and proper constraints, the dual Kalman filter method and the joint extended Kalman filter method 
are employed to simultaneously estimate the system state and the pump power. Numerical examples demonstrate the effectiveness of the proposed algorithms.
\end{abstract}

\section{INTRODUCTION}

Quantum estimation is at the heart of many research areas including quantum control, quantum computation and quantum metrology \cite{nielsen2000computation,wiseman2010measurement,Jrshin2006control}. In quantum state estimation, the aim is to estimate the state of a quantum system given measurement data. Various studies have been presented on both static state estimation and the tracking of a dynamical quantum state \cite{nielsen2000computation,wiseman2010measurement}. For a quantum state estimation problem, we usually assume that the system is well modeled. However, disturbances due to environmental fluctuations or experimental settings may lead to inaccuracies in system modeling \cite{Dong2019learning}. Thus, various studies have been done on parameter estimation which aims to estimate unknown parameters in the system modeling from obtained information \cite{xue2019spin,tsang2009optimal,yuanlong2019Automatica,yuanlong2018TAC,yuanlong2018,shu2017identify}. Further studies considered that both the state and parameters in a system model should be estimated simultaneously, motivated by either the need to estimate the system state robustly or to estimate the parameters of interest. Simultaneous state-parameter estimation has potential applications in modeling, identification and prediction \cite{ Nelson1967,connor1994,wan2001,owen1996,hossein2010,Bee2021joint}. 

In quantum research, the simultaneous state-parameter estimation problem may also have potential applications in the detection of a classical field by using a quantum sensor and in robust quantum state estimation \cite{Degen2017,Qi2020Cyber,Qi2019TCST,Shi2017Fault,qing2020fault,qing2016fault,Gao2016Fault}. A series of works on the state-parameter estimation of a quantum system have already been presented \cite{Qi2020Cyber,Qi2019TCST,qing2020fault,Shi2017Fault,qing2016fault,Gao2016Fault}. For example, in \cite{Qi2020Cyber,Qi2019TCST}, the authors modeled the unknown parameter by using a quantum analog system. The works in \cite{qing2020fault,qing2016fault,Gao2016Fault} employed the quantum-classical  Bayesian  inference  method to  solve  fault-tolerant quantum estimation problems. 

In this paper, we employ classical filtering theory for the state-parameter estimation of a quantum system to provide more flexibility on the choice of filtering methods. We consider a nondegenerate optical parametric oscillator (OPO) system which is one of the most interesting and widely used devices in quantum optics \cite{HansThird}. The pump power, a key parameter of an OPO system, may fluctuate due to two influences: unwanted disturbances from environmental noise; or by design the pump power is subject to an external signal of interest \cite{Dong2019learning,Degen2017}. In classical filtering theory, several algorithms have been proposed for simultaneous state-parameter estimation problems. For example, the dual Kalman filter (dual-KF) method was first proposed in \cite{Nelson1967} for linear systems and then developed for nonlinear systems in \cite{connor1994,wan2001}. The joint extended Kalman filter (joint-EKF) \cite{owen1996,hossein2010} combines the quantum state and
the unknown pump power into a single joint vector. Thus,
the combined system becomes nonlinear and the extended Kalman filter (EKF) is used to linearize the system. Based on existing works \cite{owen1996,hossein2010,Nelson2000Thesis}, the joint-EKF is usually expected to be on average more economic than the dual-KF while it may suffer to the divergence problem. Moreover, the joint-EKF is often more sensitive to factors that increase the estimation error due to the approximation procedure imposed by the linearization of the EKF. 


Note that classical filtering theory can not be applied to a quantum system directly since the quantum conditional expectation can not be defined properly in a classical probability space due to the uncertainty principle. However, we can find an equivalent classical problem under the assumption that the system is linear Gaussian and with proper constraints on the classical analog system. Thus, both the dual-KF and the joint-EKF can be employed to update estimates of both the state and the unknown parameter. Our main contribution is to formulate the state-space representation for an OPO system, map the quantum filtering problem to its classical analog and demonstrate the efficacy of the two employed methods for the OPO system with a dynamic parameter. Numerical results show that both the dual-KF and the joint-EKF can achieve state-parameter estimation with better performance compared to the case where no filter algorithm is applied to the unknown parameter.

This paper is organized as follows. In Section \ref{statespace}, we formulate the system dynamics of an OPO system using state-space representation. In Section \ref{filter4both}, we first present analysis on a quantum filter and its classical analog. The dynamics of the time-varying pump power are given. Then, the dual-KF method and the joint-EKF method are employed for the simultaneous state-parameter estimation. In Section \ref{simulation}, we investigate the performance of the employed algorithms. Section \ref{conclusion} concludes the paper.
\section{state-space representation of an OPO system with Homodyne measurement}\label{statespace}

We consider an OPO system consisting of a cavity and a nonlinear medium where the nonlinear effects caused by the medium can be enhanced by the cavity \cite{HansThird}. In this section, we describe the dynamics of the OPO system using a state-space representation. 

\begin{figure}[h]
	\centering		
	\includegraphics[width=8cm]{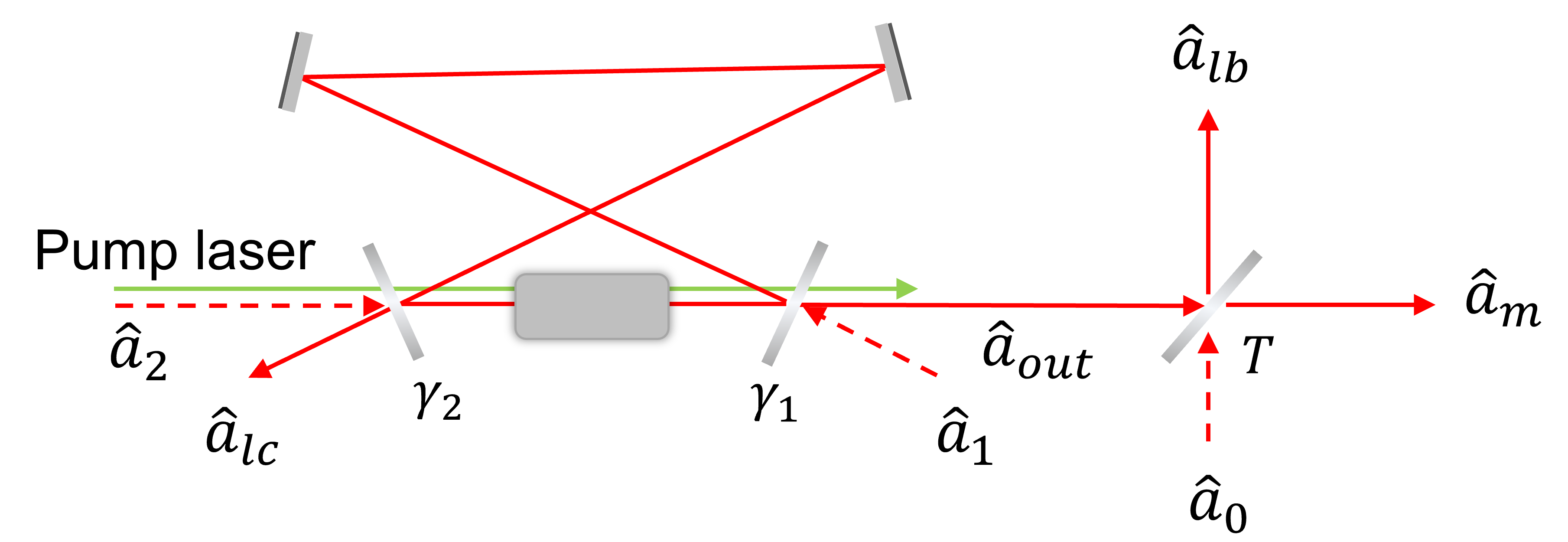}		
	\caption{Schematic of an OPO system. The cavity contains 4 mirrors and a nonlinear medium (the gray square inside the cavity). The red line is abstract representation of the cavity mode. There are two inputs $\hat{a}_{1}$ and $\hat{a}_{2}$ and the corresponding outputs $\hat{a}_{out}$ and $\hat{a}_{lc}$. The output $\hat{a}_{out}$ of the cavity is fed into a beamspiltter which yields two beams $\hat{a}_{m}$ and $\hat{a}_{lb}$. $\hat{a}_0$ is vacuum noise.}
	\label{OPOa}
\end{figure}

The cavity has two input-output channels, which is a common configuration in experiment \cite{HansThird}. The first channel with decay rate $\gamma_1$ goes to the measurement. Here, a beamsplitter is added to represent the inevitable measurement loss and noise (see Fig. \ref{OPOa}). The second channel with decay rate $\gamma_2$ is often used to model the loss inside the cavity. The total decay rate is $\gamma=(\gamma_1+\gamma_2)$. 

A nonlinear medium can be characterized by the following Hamiltonian
\begin{equation}
    H_{int}=\mi \hbar \chi (\hat{b}^\dagger \hat{a}^2-{\hat{a}^\dagger}^2\hat{b}),
\end{equation}
where $\hat{a}$ is the annihilation operator of the cavity and $\hat{b}$ is the annihilation operator of the pump beam. The coefficients $\chi$ is the second order susceptibilities of the crystal medium \cite{HansThird}. Therefore, the dynamics of the nonlinear medium is
\begin{equation}\label{nediumdynamic}
        \dot{\hat{a}} = -2 \chi \hat{a}^\dagger \hat{b}.
\end{equation}

 Under proper assumptions, one can replace the operator $\hat{b}$ in \eqref{nediumdynamic} by the $c$-number $\upepsilon$ such that
\begin{equation}
    \upepsilon=-2\chi\beta,
\end{equation}
where $\beta=\langle \hat{b} \rangle$ and $\langle \cdot \rangle$ indicates the quantum expectation \cite{bouten2007introduction}.
It can be seen that $\upepsilon$ depends on both the nonlinear medium and the pump power \cite{HansThird}. Therefore, the Hamiltonian of the OPO system can be given as 
\begin{equation}\label{HamiltonianModel1}
\begin{split}
    \hat{H}&=\hat{H}_{sys}+\hat{H}_{int},\\
    \hat{H}_{sys}&=\hbar \omega_r \hat{a}^\dagger \hat{a},\\
    \hat{H}_{int}&=\frac{1}{2}\mi\hbar[\upepsilon e^{-\mi\omega_pt}(\hat{a}^\dagger)^2 - \upepsilon^*e^{\mi\omega_pt}\hat{a}^2],\\
\end{split}
\end{equation}
where $\omega_p$ is the angular frequency of the pump laser and  $\omega_r$ is the cavity resonance angular frequency \cite{HansThird}. We assume that the pump and cavity are tuned so that $\omega_p=2\omega_r$ \cite{Collett198squeezing}. We move to a rotating frame and the dynamics of $\hat{a}$ are as follows
\begin{equation}
\begin{split}
        d\hat{a} &= (\upepsilon \hat{a}^\dagger - \gamma \hat{a} ) dt+ \sqrt{2\gamma_1}d\hat{\mathbb{A}}_{1} + \sqrt{2\gamma_2}d\hat{\mathbb{A}}_{2},\\
        d\hat{a}^\dagger &= (\upepsilon^* \hat{a} - \gamma \hat{a}^\dagger ) dt+ \sqrt{2\gamma_1}d\hat{\mathbb{A}}^\dagger_{1} + \sqrt{2\gamma_2}d\hat{\mathbb{A}}^\dagger_{2}.
\end{split}
\end{equation}
Here, we use the corresponding differential form to represent the noises (e.g., $d\hat{\mathbb{A}}_{1}=\hat{a}_{1}dt$). In this paper, we are only interested in the amplitude of the pump power. Thus, $\upepsilon$ can be regarded as a real number.

The system consisting of the OPO and the beamsplitter can be regarded as a 3-input-3-output system with corresponding inputs $\hat{a}_{0},\hat{a}_{1},\hat{a}_{2}$ and outputs $\hat{a}_{m},\hat{a}_{lb},\hat{a}_{lc}$. The input-output relation of the system is
\begin{equation}\label{in-out-cavity}
\begin{split}
        \hat{a}_{out} &= \sqrt{2\gamma_1}  \hat{a}- \hat{a}_{1}, \\
        \hat{a}_{lc} &= \sqrt{2\gamma_2}  \hat{a}- \hat{a}_{2}, \\
        \hat{a}_{m}&= \sqrt{T}\hat{a}_{out} - \sqrt{1-T}\hat{a} _{0}.
\end{split}
\end{equation}
where $T\in[0,1]$ is the transmittance of the beam splitter, which corresponds to the measurement efficiency. Let $\hat{q}$ and $\hat{p}$ denote the quadratures 
\begin{equation}
        \hat{q} \equiv \sqrt{\frac{\hbar}{2}} \left( \hat{a} + \hat{a}^\dagger  \right), \
     \hat{p} \equiv -\mi \sqrt{\frac{\hbar}{2}} \left( \hat{a}- \hat{a}^\dagger  \right).
\end{equation}
Then, we have the following output quadratures
\begin{equation}
\begin{split}
      & \hat{q}_{m} = \sqrt{2T\gamma_1} \hat{q} - \sqrt{T}\hat{v}_1- \sqrt{1-T}\hat{v}_5,\\
      & \hat{p}_{m} = \sqrt{2T\gamma_1} \hat{p} - \sqrt{T}\hat{v}_2- \sqrt{1-T}\hat{v}_6,\\
      & \hat{q}_{lc} =  \sqrt{2\gamma_2} \hat{q} -\hat{v}_3, \\
      & \hat{p}_{lc} =  \sqrt{2\gamma_2} \hat{p} -\hat{v}_4,
\end{split}
\end{equation}
where
\begin{equation*}
\begin{split}
    d\hat{v}_1 &=\sqrt{\frac{\hbar}{2}}(d\hat{\mathbb{A}}_{1}+d\hat{\mathbb{A}}^\dagger_{1}), \\
    d\hat{v}_2 &=-\mi\sqrt{\frac{\hbar}{2}}(d\hat{\mathbb{A}}_{1}-d\hat{\mathbb{A}}^\dagger_{1}),\\
    d\hat{v}_3 &=\sqrt{\frac{\hbar}{2}}(d\hat{\mathbb{A}}_{2}+d\hat{\mathbb{A}}^\dagger_{2}), \\
    d\hat{v}_4 &=-\mi\sqrt{\frac{\hbar}{2}}(d\hat{\mathbb{A}}_{2}-d\hat{\mathbb{A}}^\dagger_{2}),\\
    d\hat{v}_5 &=\sqrt{\frac{\hbar}{2}}(\hat{\mathbb{A}}_{0}+\hat{\mathbb{A}}^\dagger_{0}), \\
    d\hat{v}_6 &=-\mi\sqrt{\frac{\hbar}{2}}(\hat{\mathbb{A}}_{0}-\hat{\mathbb{A}}^\dagger_{0}).
\end{split}
\end{equation*}

Assume that the following Homodyne measurement is applied to the output $\hat{a}_{m}$,
\begin{equation}\label{measure}
\begin{split}
    \hat{y} =&\sqrt{\frac{2}{\hbar}}(\hat{q}_{m}\cos\theta_m  + \hat{p}_{m}\sin\theta_m )\\
    =& \sqrt{\frac{2}{\hbar}}(\sqrt{2T\gamma_1} \hat{q}\cos\theta_m  + \sqrt{2T\gamma_1} \hat{p}\sin\theta_m  -\sqrt{1-T}\hat{v}_5 \cos\theta_m  \\
    &- \sqrt{1-T}\hat{v}_6\sin\theta_m - \sqrt{T}\hat{v}_1\cos\theta_m - \sqrt{T}\hat{v}_2\sin\theta_m),
\end{split} 
\end{equation}
where $\theta_m$ is the phase of the Homodyne measurement.

Let $\hat{x}=(\hat{q}, \hat{p})^T$ denote the system state.
The dynamics of $\hat{x}$ and the measurement $\hat{y}$ can be described by the following state-space equations,
\begin{equation}\label{xydynamics}
\begin{split}
    d\hat{x}   &= A\hat{x} dt + Bd\hat{v}, \\
    \hat{y}dt  &= C\hat{x} dt + Md\hat{v},
\end{split}
\end{equation}
where
\begin{equation}\label{ABCM}
\begin{split}
       A  &= \begin{pmatrix} 
       \upepsilon-\gamma & 0\\
      0 & -\upepsilon-\gamma
       \end{pmatrix}, \\
       B & = \sqrt{\hbar}\begin{pmatrix} \sqrt{\gamma_1} & 0 & \sqrt{\gamma_2} & 0  & 0 & 0 \\  0 & \sqrt{\gamma_1} & 0 & \sqrt{\gamma_2}  & 0 & 0  \end{pmatrix},\\
       C & = 2\sqrt{\frac{T\gamma_1}{\hbar}} \begin{pmatrix} 
       \cos{\theta_m} & \sin{\theta_m} 
       \end{pmatrix},\\
       M & = -\sqrt{\frac{2}{\hbar}}\begin{pmatrix}
        \sqrt{T}\cos{\theta_m} \\ \sqrt{T}\sin{\theta_m} \\ 0 \\ 0 \\ \sqrt{1-T}\cos{\theta_m} \\ \sqrt{1-T}\sin{\theta_m} 
       \end{pmatrix}^T,\\
       \hat{v} &=  ( \hat{v}_1\ \hat{v}_2\ \hat{v}_3\ \hat{v}_4\ \hat{v}_5\ \hat{v}_6 )^T.
\end{split}
\end{equation}
The quantum covariance of two operator vectors $\hat{o}_1$ and $\hat{o}_2$ is defined as
\begin{equation}\label{variancedef}
    Cov(\hat{o}_1,\hat{o}_2)\equiv \frac{1}{2}\langle \hat{o}_1\hat{o}_2^T+(\hat{o}_2\hat{o}_1^T)^T\rangle-\langle \hat{o}_1\rangle \langle \hat{o}_2\rangle^T.
\end{equation}
 Denote the correlation matrices as 
\begin{equation}\label{variances}
    \begin{split}
        Ddt &=Cov(Bd\hat{v},Bd\hat{v}),  \\
        \Gamma^Tdt &= Cov(Bd\hat{v},Md\hat{v}), \\
        Rdt &= Cov(Md\hat{v},Md\hat{v}).
    \end{split}
\end{equation}
Let $V$ denote the covariance matrix $V=Cov(\hat{x},\hat{x})$.
Then, the Heisenberg uncertainty principle \cite{Robertson1929} gives
\begin{equation}\label{SUncertain}
    \det(V)\geq \frac{1}{4}|\langle[\hat{q},\hat{p}]\rangle|^2.
\end{equation}
Since the commutation relation of quadratures gives $[\hat{q},\hat{p}]=\mi\hbar$, the uncertainty principle \eqref{SUncertain} reads
\begin{equation}\label{SUncertain02}
    \det(V)\geq \frac{\hbar^2}{4}
\end{equation}
which can be rewritten as
\begin{equation}\label{HUncertain02}
    V +\frac{\mi \hbar\Sigma}{2} \geq 0
\end{equation}
    where $\Sigma_{kl}=-\mi [\hat{x}_k,\hat{x}_l]$ is called the symplectic matrix \cite{wiseman2010measurement}. In our case, the symplectic matrix is
\begin{equation}
    \Sigma=\begin{pmatrix}0&1\\-1 &0\end{pmatrix}
\end{equation}
which yields
\begin{equation}
    V+ \hbar \begin{pmatrix}0&\frac{\mi}{2}\\\frac{-\mi}{2} &0\end{pmatrix}\geq 0.
\end{equation}
\section{Filter for both the state and the time-varying pump power}\label{filter4both}

In this section, we consider the situation where the pump power $\upepsilon$ is a random process. To achieve the simultaneous state-parameter estimation, we first provide standard quantum filter for the system state. Then, we combine the state filter and the classical parameter filter after an explanation of the existence of a classical analog to our quantum filtering problem.

Given the dynamics of the OPO system in \eqref{xydynamics}, a filtered quantum state conditioned on a measurement record can be obtained by using quantum filtering theory \cite{bouten2007introduction}. However, the estimation of the classical pump power and the quantum state can not be unified due to the difference between quantum and classical mechanics. Fortunately, there exists a classical analog of the quantum filtering equations given that the quantum system is linear Gaussian \cite{wiseman2010measurement,Doherty2000control,Wiseman2000unravelling}. We first provide a treatment on the equivalence of a quantum conditioned state and its classical analog. Then, the dual-KF method and joint-EKF method are employed to solve the simultaneous state-parameter filtering problem based on the quantum-classical equivalence.

\subsection{Filter for the state}
Given the system Hamiltonian in \eqref{HamiltonianModel1} and the measurement in \eqref{measure}, the conditioned dynamics of the system can be obtained by using the standard quantum filtering theory \cite{bouten2007introduction}. Meanwhile, a quantum state can also be characterized by a Wigner distribution which is a pseudo-probability distribution in the phase space
over a classical configuration corresponding to the quantum configuration \cite{wiseman2010measurement}. The Wigner function appears like a joint classical probability distribution in classical cases. A quantum system can be described as a linear Gaussian system if its Wigner function is Gaussian and dynamics are linear. The Wigner function is positive-definite for a Gaussian quantum state while it can be negative for general quantum states. For a linear quantum system, the Gaussianity can be preserved since future states are linear combinations of the initial Gaussian state.  Therefore, the dynamics of the system \eqref{xydynamics} are completely described by the time evolution of the first and second statistical moments of the quadrature coordinates \cite{wiseman2010measurement,Doherty1999feedback,Doherty2000Quantum,Wiseman2000unravelling}. In our case, the moments of the unconditioned Gaussian state are
\begin{equation}\label{LGQsystem}
\begin{split}
    d\langle \hat{x}\rangle   &= A\langle \hat{x}\rangle dt, \\
    \frac{dV}{dt} &=AV+VA+D.
\end{split}
\end{equation}
Since the state-space equations \eqref{xydynamics} are linear, with the assumption that the initial state is Gaussian, the conditioned state remains a Gaussian state with the following moments
\begin{equation}\label{filterforstate}
    \begin{split}
    d\langle \hat{x}\rangle_c   &= A\langle \hat{x}\rangle_c dt + (V_cC^T+\Gamma^T)R^{-1}dw, \\
    \frac{dV_c}{dt} &=AV_c+V_cA^T+D-(V_cC^T+\Gamma^T)R^{-1}(CV_c+\Gamma),
    \end{split}
\end{equation}
where $dw=ydt-C\langle \hat{x}\rangle_c dt$ is the innovation. Here, the subscript $c$ means the quantity is conditioned on a measurement record.

 Note that both \eqref{xydynamics} and \eqref{filterforstate} are isomorphic to those of a classical linear Gaussian system. Then, we can find the following classical system analog to the system \eqref{xydynamics} \cite{wiseman2010measurement}
\begin{equation}\label{xyclassical}
\begin{split}
    dx   &= Axdt + Bdv, \\
    ydt  &= Cxdt + Mdv,
\end{split}
\end{equation}
where $x=(q,p)^T$ is a vector of classical random variables and $y$ is a classical random variable. $v$ is a classical Wiener process. The coefficient matrices are given in \eqref{ABCM}. 

Here, we present the main restrictions applied to the classical system \eqref{xyclassical} inherited from the quantum origin. The two restrictions are originated from the unitary evolution of the unconditioned system and the Heisenberg uncertainty principle. For the unconditioned system \eqref{xyclassical}, the unitarity places the following fluctuation-dissipation restriction on the drift and diffusion matrices $A$ and $D$ \cite{wiseman2010measurement}
\begin{equation}
    D-\mi\hbar(A\Sigma-\Sigma^TA^T)/2\geq 0.
\end{equation}
For the conditioned state, we have the following fluctuation-observation relation which preserves the uncertainty relation \eqref{HUncertain02}
\begin{equation}
    D-\Gamma^T\Gamma-\frac{\hbar^2}{4}\Sigma C^TC\Sigma^T\geq 0.
\end{equation}
Since we assume that the inputs are vacuum states, which are Gaussian, the covariance for the vacuum noises is $\frac{\hbar}{2}I$ where $I$ is the identity. The Heisenberg uncertainty relation for the noise vector $v$ gives \cite{Ferraro2005Gaussian}
\begin{equation}
    Cov(v,v) +\frac{\mi \hbar\Omega}{2} \geq 0
\end{equation}
where 
\begin{equation}
    \Omega=\oplus_{k=1}^3 \Sigma=\begin{pmatrix} \Sigma & & \\
                & \Sigma & \\
                & & \Sigma \end{pmatrix}.
\end{equation}

\subsection{Time varying pump power}\label{pumppower}

In optical cases, the pump power $\upepsilon$ may fluctuate due to environmental perturbations or instrumental settings. Here, we assume that $\upepsilon$ is described by the following stochastic process  
\begin{equation}\label{epsilondynamic}
    d\upepsilon=\mu (\upepsilon-c) dt + g dv_\upepsilon
\end{equation}
where $\mu<0$ is the drift coefficient and $g$ is the diffusion coefficient. $v_\upepsilon$ is a classical Wiener process. The constant $c$ characterizes the expected value of the stochastic process and we denote it as the tendency constant. 
Thus, the corresponding state-space equations are
\begin{equation}\label{pumpandmeasure}
    \begin{split}
        d\upepsilon &=\mu (\upepsilon-c) dt +gdv_\upepsilon, \\
         ydt  &= Cx dt + Mdv.\\
    \end{split}
\end{equation}
The covariance of two vectors $o_1$ and $o_2$ of classical random variables is defined as
\begin{equation}
    Cov(o_1,o_2) \equiv \mathbb{E}(o_1o_2^T)-\mathbb{E}(o_1)\mathbb{E}(o_2^T),
\end{equation}
where $\mathbb{E}(\cdot)$ denotes a classical expectation.
The correlation matrices are
\begin{equation}
    \begin{split}
        D_\upepsilon dt &=Cov(gdv_\upepsilon,gdv_\upepsilon),  \\
        \Gamma_\upepsilon^Tdt &= Cov(gdv_\upepsilon,Mdv), \\
        Rdt &= Cov(Mdv,Mdv).
    \end{split}
\end{equation}
\subsection{The dual-KF method}\label{dualestimation}
 In this section, we briefly revisit the dual-KF method from the Bayes estimation perspective and provide the continuous dual-KF algorithm.
Given \eqref{xyclassical} and \eqref{pumpandmeasure}, we aim to obtain estimates of both $\upepsilon$ and the state $x$. The task can be achieved by using the Maximum a Posteriori (MAP) method which aims to maximize the joint conditional density $\rho_{x,\upepsilon|y}$ \cite{Nelson2000Thesis,wan2001}. The dual-KF method separates the conditional density into two terms
\begin{equation}
    \rho_{x,\upepsilon|y}=\rho_{x|\upepsilon,y}\rho_{\upepsilon|y}.
\end{equation}
The estimate $\upepsilon_c$ is found by maximizing $\rho_{\upepsilon|y}$ while the estimate $x_c$ is found by maximizing $\rho_{x|\upepsilon_c,y}$. 

The corresponding cost function for the state $x$ is 
\begin{equation}\label{xcost}
    J(x,\upepsilon_c)= \int \left(\frac{(y-Cx)^2}{R} + \frac{(x-x_c^-)^2}{D}  \right)dt
\end{equation}
 where $x_c^-=E[x | y_0^{t-dt},\upepsilon_0^{t-dt}]$ is the optimal prediction. Since the probability density is Gaussian, minimization of the above cost function regarding $x$ can be reached by using the Kalman-Bucy filter \eqref{filterforstate}.

Similarly, the cost function for the estimation of parameter $\upepsilon$ is

\begin{equation}\label{epsiloncost}
    J(\upepsilon,x_c)= \int \left(\frac{(y-y^-)^2}{R} + \frac{(\upepsilon-\upepsilon_c^-)^2}{D_\upepsilon}  \right)dt
\end{equation}
 where $y^-=E[y | y_0^{t-dt},\upepsilon_0^{t-dt}]$ and $\upepsilon_c^-$ are the optimal predictions. The cost can be minimized using the following Kalman-Bucy filter:
\begin{equation}\label{filterforpump}
    \begin{split}
        d\upepsilon_c &= \mu (\upepsilon_c-c) dt + K_\upepsilon d\eta , \\
        K_\upepsilon &=(V_{c,\upepsilon}C_\upepsilon^T+\Gamma_\upepsilon^T) R^{-1},  \\
        \frac{dV_{c,\upepsilon}}{dt} &= \mu  V_{c,\upepsilon}+V_{c,\upepsilon} \mu +D_\upepsilon-K_\upepsilon R K^T_\upepsilon,
    \end{split}
\end{equation}
where $d\eta=ydt-Cx_c dt$ is the innovation and
\begin{equation}
    \begin{split}
       C_\upepsilon &=  \frac{\partial (Cx)}{\partial \upepsilon}=C\frac{\partial x}{\partial \upepsilon}= C\frac{\partial A}{\partial \upepsilon}x
    \end{split}
\end{equation}
is the linearization coefficient which can be calculated using \eqref{filterforstate} while the conditioned $x_c$ is used to approximate $x$. At every time step, the conditioned state $x_c$ is updated using the current estimated parameter $\upepsilon_c$ while the estimated $\upepsilon_c$ is also updated using the current state $x_c$.
Here, we summarize the algorithm as follows.

\begin{algorithm} (continuous dual-KF)

Initialized with:

$\upepsilon_c=\upepsilon_0$, $V_{c,\upepsilon}=\mathbb{E}[(\upepsilon_0-\upepsilon_c)(\upepsilon_0-\upepsilon_c)^T]$,

$x_c=x_0$, $V_c=\mathbb{E}[(x_0-x_c)(x_0-x_c)^T]$.

Update for the state:
    
 $K_x=(V_cC^T+\Gamma^T)/R$,
    
 $dx_c= Ax_c dt + K_xdw$,  
    
 $\frac{dV_c}{dt}= AV_c+V_cA^T+D-K_xRK_x^T$. 

Update for the pump power:

$d\upepsilon_c = \mu (\upepsilon_c-c) dt + K_\upepsilon d\eta$ ,
 
$K_\upepsilon =(V_{c,\upepsilon}C_\upepsilon^T+\Gamma_\upepsilon^T) R^{-1}$,  
        
$\frac{dV_{c,\upepsilon}}{dt} = \mu  V_{c,\upepsilon}+V_{c,\upepsilon} \mu +D_\upepsilon-K_\upepsilon R K^T_\upepsilon$.
\end{algorithm}

\subsection{The joint-EKF method}\label{joint-EKF}
The joint estimation aims to maximize the following conditional density
\begin{equation}
    \{x_c,\upepsilon_c\}=\argmax_{x,\upepsilon} \rho_{x,\upepsilon|y}.
\end{equation}
To generate the above MAP estimates, we define the following new joint state consisting of both $x$ and $\upepsilon$
\begin{equation}
    z=\begin{pmatrix} q & p & \upepsilon-c \end{pmatrix}^T.
\end{equation}
The state-space representation of the joint state $z$ is
\begin{equation}\label{zdynamics}
    \begin{split}
        dz&=A_zzdt + B_zdv_z,\\
        y_zdt & =C_zzdt+M^zdv_z,
    \end{split}
\end{equation}
where
\begin{equation}
    \begin{split}
        A_z &=\begin{pmatrix} \upepsilon-\gamma & 0 & 0 \\ 0 & -\upepsilon-\gamma & 0 \\ 0 & 0& \mu \end{pmatrix}, \\
        B_z &=\begin{pmatrix} \sqrt{\hbar \gamma_1} & 0 & \sqrt{\hbar \gamma_2} & 0 & 0 & 0 & 0 \\ 0& \sqrt{\hbar \gamma_1} & 0 & \sqrt{\hbar \gamma_2} & 0  & 0 & 0  \\ 0 & 0& 0 &0 &0 &0 &g\end{pmatrix}, \\
        C_z &=2\sqrt{\frac{T\gamma_1}{\hbar}}\begin{pmatrix} \cos{\theta_m} & \sin{\theta_m} & 0  \end{pmatrix}, \\
        M_z &= -\sqrt{\frac{2}{\hbar}}\begin{pmatrix} \sqrt{T}\cos{\theta_m} \\ \sqrt{T}\sin{\theta_m} \\ 0 \\ 0 \\ \sqrt{1-T}\cos{\theta_m} \\ \sqrt{1-T}\sin{\theta_m} \\ 0\end{pmatrix}^T, \\
        v_z &= \begin{pmatrix} v_x\\v_\upepsilon  \end{pmatrix}.
    \end{split}
\end{equation}
The noise correlation matrices are
\begin{equation}
    Cov\begin{pmatrix} B_zdv_z \\ M_zdv_z \end{pmatrix} \begin{pmatrix} (B_zdv_z)^T & (M_zdv_z)^T \end{pmatrix} =\begin{pmatrix} R^z & R^{zy} \\ (R^{zy})^T & R^{y_z}\end{pmatrix}dt.
\end{equation}

The generalized system becomes nonlinear and an EKF can be applied to generate an approximation of the MAP estimation. We denote this method as the joint-EKF method and the algorithm is as follows.
\begin{algorithm} (joint-EKF) 

Initialized with:

$z_c=z_0$, 

$V_0=\mathbb{E}[(z_0-z_c)(z_0-z_c)^T]$.

Update for the state:

$K_z=(V_cC_z^T+\Gamma^T)/R^{y_z}$,
    
$dz_c=  A_{z,c}z_c dt + K_zdw$,  
    
$\frac{dV_c}{dt}=\bar{A}_{z,c}V_c+V_c\bar{A}_{z,c}^T+R^z-K_zR^{y_z}K_z^T$,

where $A_{z,c}=A_{z}|_{z=z_c}$ is an approximate of $A_z$ given current estimate of the state $z_c$ and $\bar{A}_{z,c}=\frac{\partial A_z z}{\partial z}|_{z=z_c}$.
\end{algorithm}

\section{Numerical results}\label{simulation}
In this section, we first apply the proposed algorithms to a special case where parameters are fixed to values that are close to real experiments. Here, we aim to show the effectiveness of the two methods without comparing their performance. Then we test the performance improvement regarding the measurement efficiency, the tendency constant and diffusion coefficient of the classical signal.
\subsection{Case study}\label{case study}
In this section, simulation results of both the dual-KF method and the joint-EKF method are presented. For comparison, we simulate $x_c$ using the dual-KF method, the joint-EKF method and the KF method. For the KF method, the Kalman filter is applied to the system state while no filter is applied to $\upepsilon$ and we assume that $\upepsilon=c$ is a constant. $x_T$ represents the true evolution of the system state. Here, $x_T$ is simulated using \eqref{filterforstate} given known $\upepsilon$ and a complete measurement record of all three outputs $\hat{a}_{m}$, $\hat{a}_{lb}$ and $\hat{a}_{lc}$. See the Appendix for a detailed explanation of the simulation of $x_T$.

To quantify the performance improvement, we define the \textit{relative performance improvement} (RPI) for $\upepsilon$ as
\begin{equation}\label{normImprove}
    \mathcal{I}_{\upepsilon_c}=1-\frac{\int (\upepsilon_c-\upepsilon_T)^2dt}{\int (\upepsilon_{KF}-\upepsilon_T)^2dt}.
\end{equation}
The subscript $c$ indicates either the dual-KF method or the joint-EKF method. The RPI of $\upepsilon_c$ evaluates the percentage of the MSE reduced by using the two proposed methods compared with the KF method. We simulate the system for $N$ times and characterize the RPI using its mean value $\mathcal{I}^m_{\upepsilon_c}$ and the \textit{standard error of the mean} (SEM) $S$ which are defined as follows
\begin{equation}\label{normImprove02}
\begin{split}
    \mathcal{I}^m_{\upepsilon_c} &=\frac{1}{N}\sum_{j=1}^N \mathcal{I}_{\upepsilon_c,j},\\
    S &=\sqrt{\frac{\sum(\mathcal{I}_{\upepsilon_c,j}-\mathcal{I}^m_{\upepsilon_c})^2}{N(N-1)}}.
\end{split}
\end{equation}
The relative improvement for the quantum state $\mathcal{I}_{x_c}=(\mathcal{I}_{q_c};\mathcal{I}_{p_c})$ is defined similarly and the corresponding mean and SEM are calculated in the same way.

The following parameters are selected for the case study.
\begin{itemize}
    \item $\hbar =1$,
    \item $x_0=(0 \  0)$,
    \item $T= 1$,
    \item $\theta_{m}=\frac{\pi}{12}$~rad, 
    \item $\gamma_1= 0.95$~rad/s,\ $\gamma_2 = 0.05$~rad/s, 
    \item $c=0.5,\ \mu=-0.01$~rad/s, $g=0.028$.
\end{itemize}
 Note that the escape efficiency $\gamma_1 / \gamma$ is based on experiments \cite{suzuki20067db}. We use the total decay rate as $\gamma = \gamma_1 + \gamma_2 = 1$~rad/s for simplicity, which determines the scaling of time evolution. The actual values in the experiments are in order of $2\pi\times10\times10^6$~rad/s. 
   
 \begin{figure}	
	\centering		
	\includegraphics[width=9.5cm]{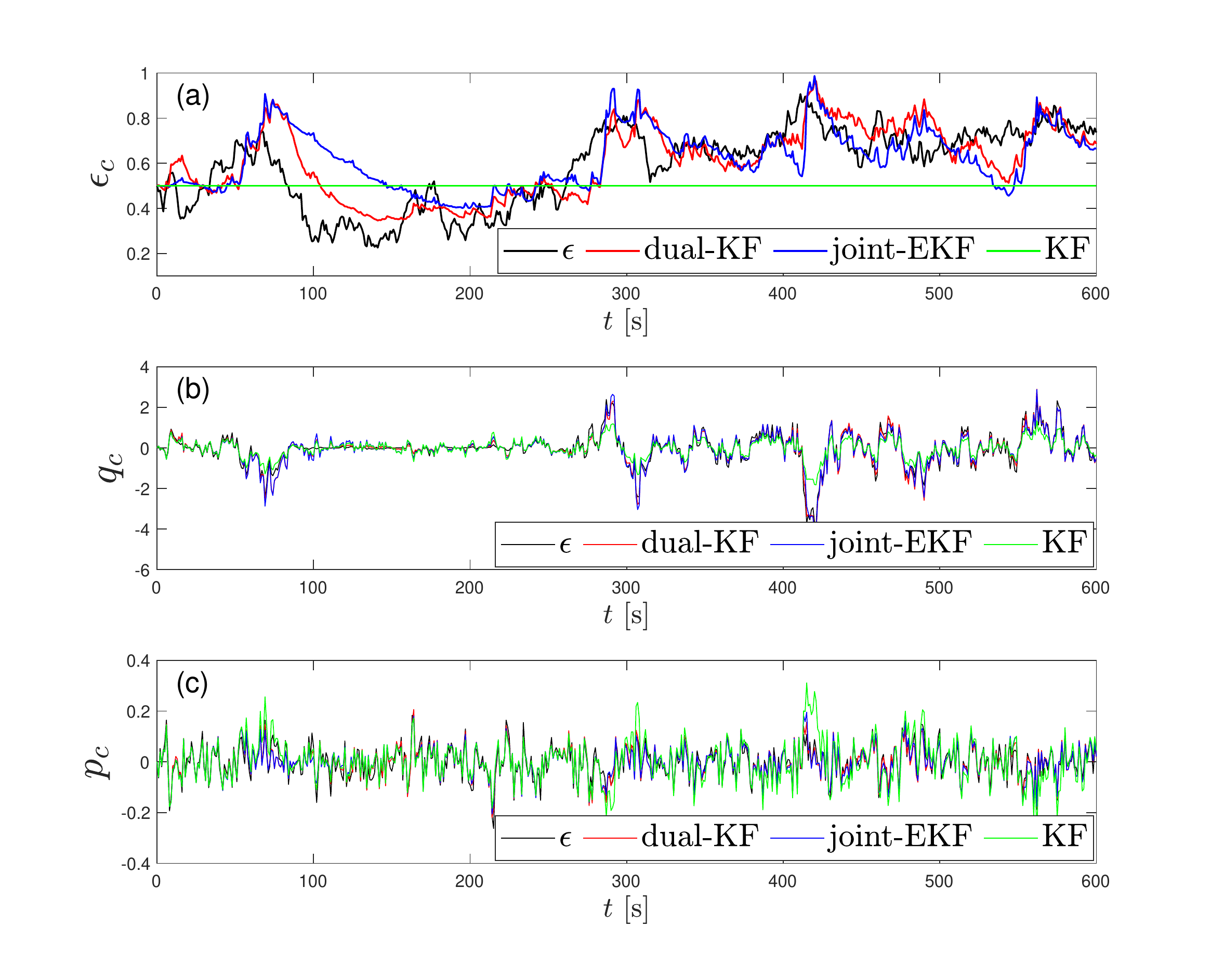}
	\caption{ One estimation trial of (a) $\upepsilon_c$, (b) $q_c$ and (c) $p_c$ by using the dual-KF method, the joint-EKF method and the KF method. }
	\label{epsilon_x00}
\end{figure}
 
 Fig. \ref{epsilon_x00} presents time evolution of one trial of the simulation.  Fig. \ref{epsilon_x00} (a) gives the time evolution of the estimated $\upepsilon_c$. The black line is the true evolution of $\upepsilon$. The green line is straight which means that $\upepsilon$ is a constant value since there is no filter applied to $\upepsilon$ for the KF method. The red and blue lines are generated by the dual-KF method and the joint-EKF method, respectively. It can be seen that on average the blue and red lines are closer to the black line compared with the green line, especially when there is a large deviation of the real value of $\upepsilon$ to the tendency constant $c$. Fig. \ref{epsilon_x00} (b) and (c) give time evolution of $q$ and $p$ of the dual-KF, the joint-EKF and the KF methods, respectively. The red line is generated by using the dual-KF method and the blue line is  generated by using the joint-EKF method. The KF method (the green line) does not take any information from measurement to update $\upepsilon_c$. The RPIs for the dual-KF method are $73.4\%$ ($\upepsilon_c$), $79.9\%$ ($q_c$) and $75.1\%$ ($p_c$) while the RPIs of the joint-EKF method are  $48.6\%$ ($\upepsilon_c$), $71.8\%$ ($q_c$) and $69.5\%$ ($p_c$). It is clear that both the proposed two methods can follow the true evolution of the system state $x$ better than the KF method.
 
 \begin{table}[H]
\begin{tabular}{|c|c|c|}
\hline
 & dual-KF &  joint-EKF\\ \hline
 $\mathcal{I}^m_{\upepsilon_c}$ (\%) &  48.6 & 38.5 \\ \hline
 $\mathcal{I}^m_{q_c}$ (\%)          &  51.1  & 42.9   \\ \hline
 $\mathcal{I}^m_{p_c}$ (\%)          &  39.5 &  34.9\\ \hline
\end{tabular}
\caption{Mean RPI of the dual-KF method and the joint-EKF method. The corresponding SEM is at most $0.92\%$.}\label{table0}
\end{table}
 
  While the performance improvement of the proposed methods may not be convincing with only one trial result, the superiority can be confirmed by Table \ref{table0} which gives the mean RPI of the dual-KF and joint-EKF methods calculated over $N=1\times 10^3$ trials. The result shows that both the dual-KF method and the joint-EKF method have a performance improvement of over $34.9\%$ while the dual-KF method provides higher RPIs on average. We also consider different initial states for both the vacuum state and the coherent state including $x_0=(0\ 0)$, $x_0=(1.4\ 0)$, $x_0=(0\ 1.4)$ and $x_0=(1.4\ 1.4)$. The RPIs are consistent for the four different initial states. Thus, we conclude that both the dual-KF method and the joint-EKF method can provide the better estimates of the system state and the pump power simultaneously than the KF method while the dual-KF method is better on average than the joint-EKF method for given system parameters for the proposed case. The superiority of the dual-KF method can be attributed to the fact that the joint-EKF method is sub-optimal since linearization is used for approximation.
 

\subsection{Performance improvement regarding varying parameters}\label{sensitivity}

In this section, we present the mean RPIs regarding the following varying parameters:
\begin{itemize}
    \item the measurement efficiency $T$,
    \item the diffusion coefficient $g$,
    \item the tendency constant $c$.
\end{itemize}
The mean RPIs and the corresponding SEMs are calculated over $N=1\times 10^3$ trials.

\begin{figure}	
	\centering		
	\includegraphics[width=9.5cm]{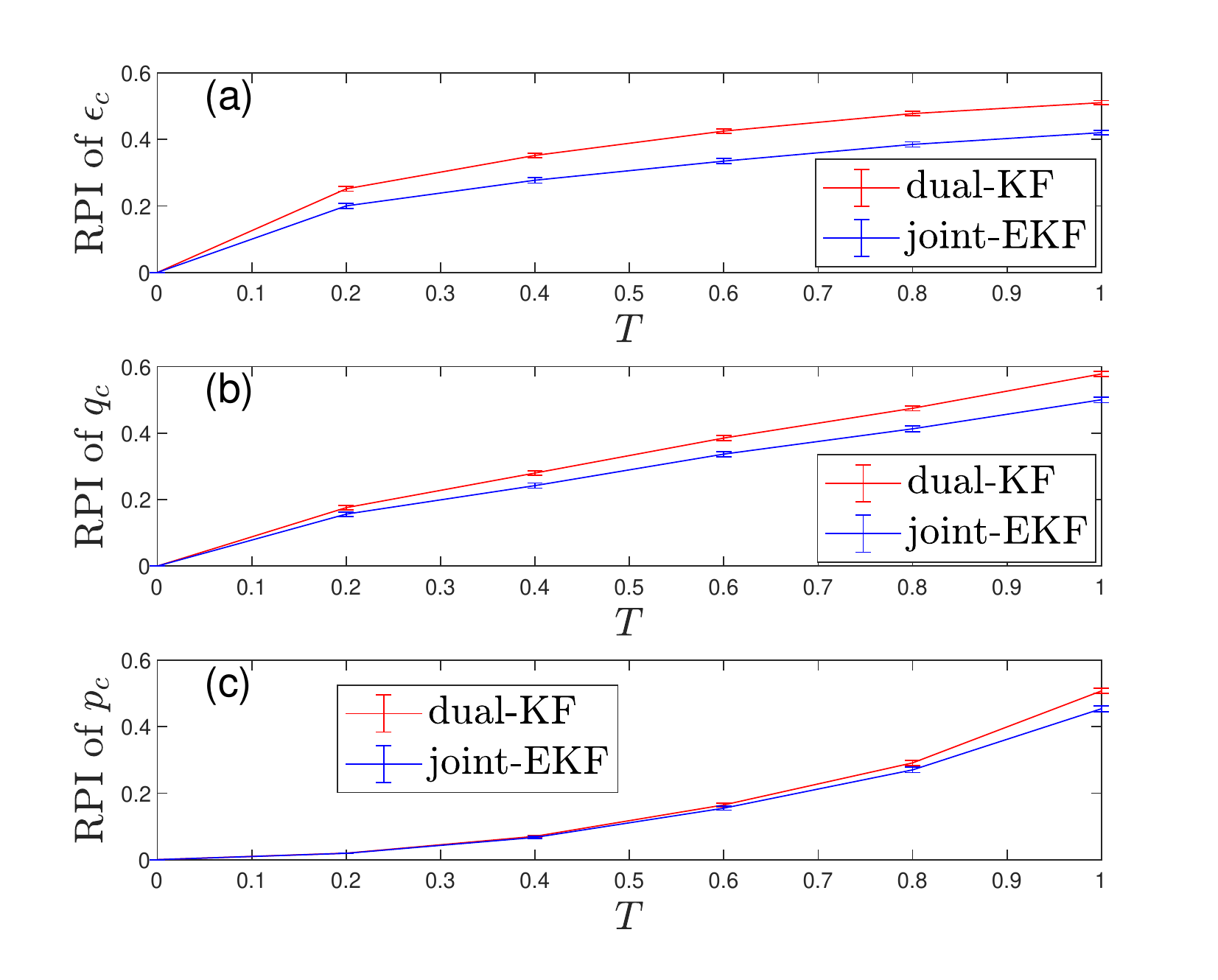}	
	\caption{RPI of (a) $\upepsilon_c$, (b) $q_c$ and (c) $p_c$ with the measurement efficiency $T$ increasing from $0$ to $1$.}
	\label{RPIVST_dev}
\end{figure}

In Fig. \ref{RPIVST_dev}, $T$  varies from $0$ to $1$ while the other parameters are given as in Section \ref{case study}. It can be seen that the mean RPI of both dual-KF method (red line) and joint-EKF method (blue line) has a positive relationship with the measurement efficiency. This is reasonable since a higher measurement efficiency means more information about the system is used for the estimation of $\upepsilon$ and a better estimated $\upepsilon$ yields better estimated state vector $x$.

\begin{figure}	
	\centering		
	\includegraphics[width=9.5cm]{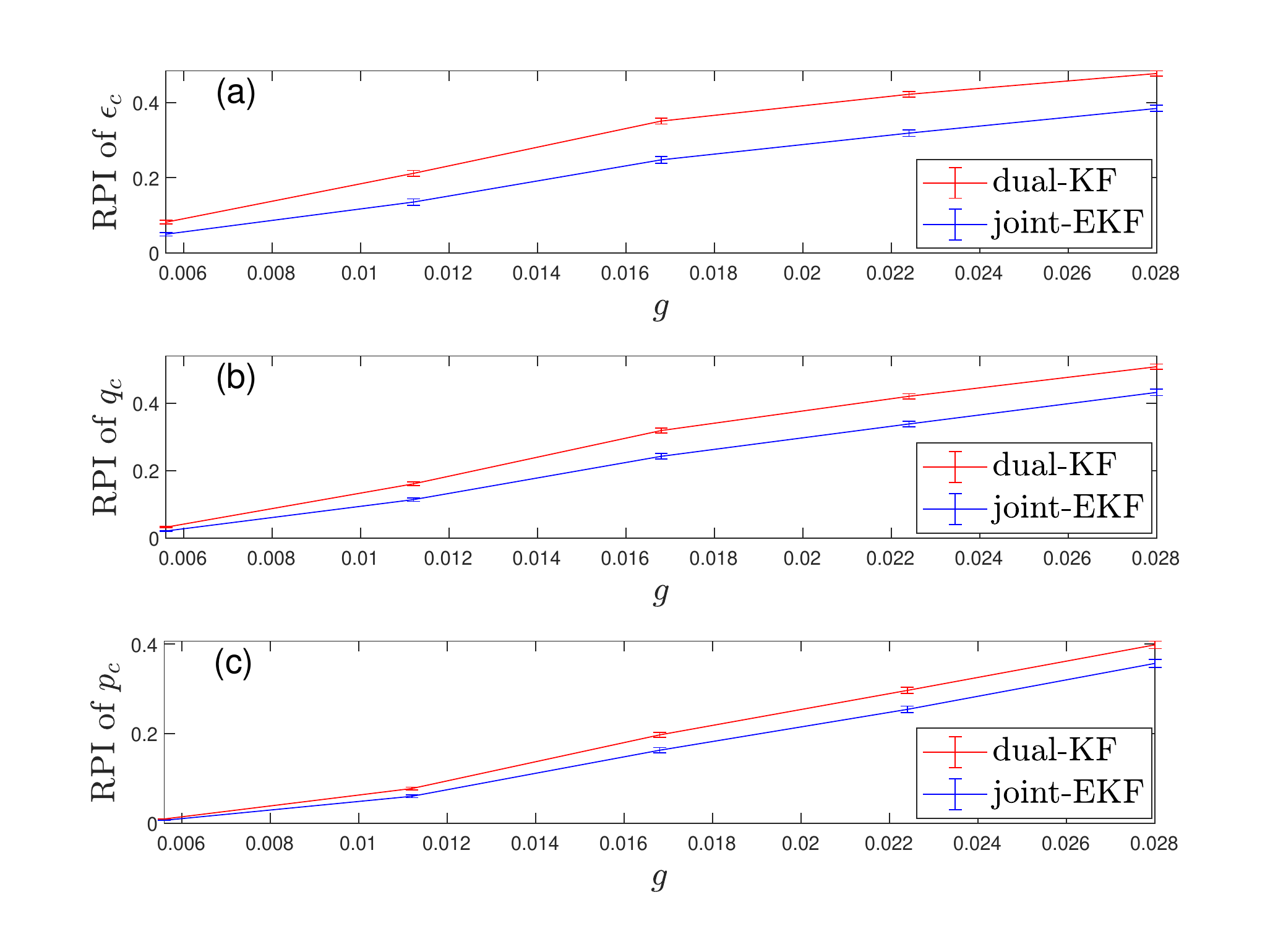}	
	\caption{RPI of (a) $\upepsilon_c$, (b) $q_c$ and (c) $p_c$ with the diffusion coefficient $g$ increasing from $0.005$ to $0.028$.}
	\label{RPIVSg}
\end{figure}

In Fig. \ref{RPIVSg}, the diffusion coefficient $g$ increases from $0.005$ to $0.035$ while the other parameters are given as in Section \ref{case study}. With the increasing of $g$, the mean RPI of $\upepsilon_c$ increases from around $0$ up to around $48\%$ (dual-KF) and $38\%$ (joint-EKF) while the mean RPI of $q_c$ and $p_c$ increase from around $0$ up to around $51\%$ (dual-KF), $43\%$ (joint-EKF) and $40\%$ (dual-KF), $36\%$ (joint-EKF), respectively. Thus, we see that the proposed algorithms show more improvement when the diffusion coefficient is larger. The reason is that a larger noise ratio results in a larger deviation of $\upepsilon$ to its tendency constant $c$ and the superiority of the proposed methods is clearer. An extreme situation is that the performance of proposed methods is the same with the KF method when the diffusion coefficient is zero.

\begin{figure}	
	\centering		
	\includegraphics[width=9.5cm]{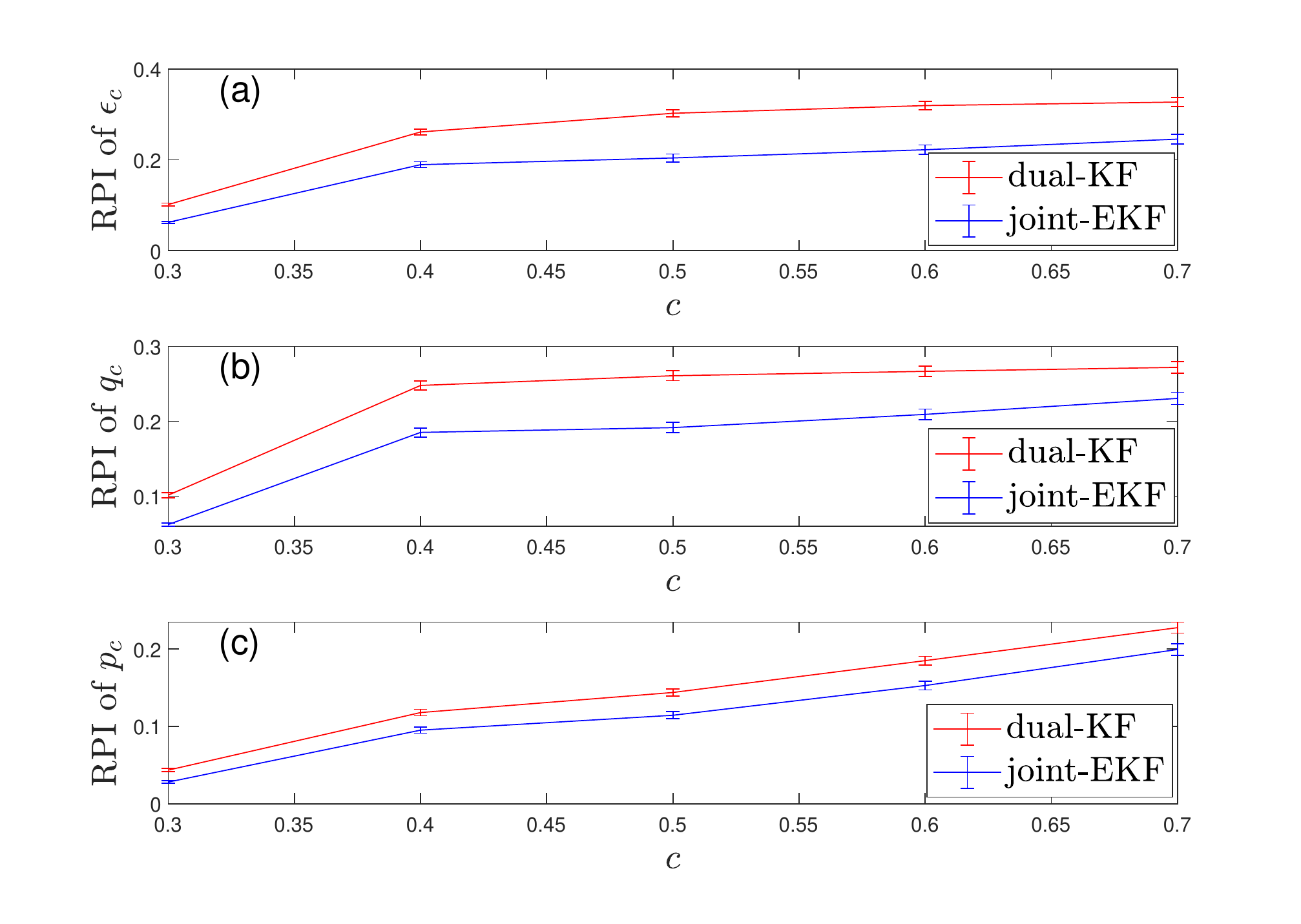}	
	\caption{RPI of (a) $\upepsilon_c$, (b) $q_c$ and (c) $p_c$ with the tendency constant $c$ increasing from $0.3$ to $0.7$.}
	\label{RPIVSc}
\end{figure}

In Fig. \ref{RPIVSc}, the tendency constant $c$ increases from $0.3$ to $0.7$ while $g=0.025$ and the other parameters are given as in Section \ref{case study}. With the increasing of $c$, the mean RPI of $\upepsilon_c$ increases from $10\%$ to $33\%$ for dual-KF and $6\%$ to $25\%$ for joint-EKF while the mean RPI of $q_c$ and $p_c$ increase from $10\%$ to $27\%$ for dual-KF of $q_c$, from $6\%$ to $23\%$ for joint-EKF of $q_c$, from $4\%$ to $23\%$ for dual-KF of $p_c$, from $3\%$ to $20\%$ for joint-EKF of $p_c$, respectively. The results show that there is a positivity relationship between the tendency constant $c$ and the RPI although the increase of RPI is small compared with the results in Fig. \ref{RPIVST_dev} and Fig. \ref{RPIVSg}. Thus, we see that the RPI is more determined by the measurement efficiency and the diffusion coefficient than the tendency constant.




\section{Conclusion}\label{conclusion}
In this paper, we considered the state and parameter estimation problem of an OPO system where the pump power is subject to a stochastic process. We first formulated the system dynamics in the state-space representation. Then, we provided analysis on quantum filtering and its classical analog under the assumption that the quantum system is linear and Gaussian. Details of restrictions on obtaining a classical analog for the quantum system were provided. Thus, the simultaneous estimation problem of both the state and the unknown parameter can be solved by using the dual-KF method and the joint-EKF method. For the dual-KF, the estimation of the state and the pump power is decoupled and two Kalman filters run concurrently. For the joint-EKF method, the state vector is augmented with the unknown pump power. Therefore, the augmented system is nonlinear and an EKF is then employed. The simulation results show that both these methods can achieve good estimation of the system. An average improvement of $30\%$-$50\%$ can be reached compared with Kalman filter method for the studied case.
The simulation results also show that the performance of the proposed algorithms increases with increasing of parameters $T$, $g$ and $c$. This work can also be extended to the state-parameter estimation of other systems. 

\addtolength{\textheight}{0cm}
\begin{appendix}
The true evolution of the system state $x_T$ is approximated by the conditioned state $x_c$ obtained by using a complete measurement record and known $\upepsilon$. We consider Homodyne measurement to all the three outputs $\hat{a}_{m}$, $\hat{a}_{lb}$ and $\hat{a}_{lc}$. Thus, measurement of $\hat{a}_{m}$ is
\begin{equation}
    \hat{y}_{\theta_m} = \sqrt{\frac{2}{\hbar}} (\cos\theta_m \hat{q}_{m} + \sin\theta_{m} \hat{p}_{m}).
\end{equation}
Measurement of $\hat{a}_{lb}$ is
\begin{equation}
     \hat{y}_{\theta_{lb}} = \sqrt{\frac{2}{\hbar}} (\cos\theta_{lb}\ \hat{q}_{lb} + \sin\theta_{lb} \hat{p}_{lb}).
\end{equation}
Measurement of $\hat{a}_{lc}$ is
\begin{equation}
   \hat{y}_{\theta_{lc}} = \sqrt{\frac{2}{\hbar}} (\cos\theta_{lc}\ \hat{q}_{lc} + \sin\theta_{lc} \hat{p}_{lc}).
\end{equation}
The complete measurement is $\hat{\tilde{y}}=(\hat{y}_{\theta_{m}},\ \hat{y}_{\theta_{lb}},\ \hat{y}_{\theta_{lc}})$. For simulation, we let $\theta_m=\theta_{lb}=\theta_{lc}$. The corresponding measurement matrices are given in \eqref{completemeasure}. Then, the system state can be estimated using \eqref{filterforstate} with the coefficient matrix given in \eqref{completemeasure} while the innovation is $dw=\hat{\tilde{y}}dt-C\langle \hat{x}\rangle dt$. 

\begin{widetext}
\begin{equation}\label{completemeasure}
\begin{split}
      C & = \frac{2}{\sqrt{\hbar}}\begin{pmatrix} 
      \sqrt{T\gamma_1} \cos{\theta_{m}} & \sqrt{T\gamma_1} \sin{\theta_{m}} \\ 
      \sqrt{\gamma_1(1-T)} \cos{\theta_{lb}} & \sqrt{\gamma_1(1-T)} \sin{\theta_{lb}} \\
      \sqrt{\gamma_2} \cos{\theta_{lc}} & \sqrt{\gamma_2} \sin{\theta_{lc}} 
      \end{pmatrix},\\
      M & =
      -\sqrt{\frac{2}{\hbar}} \begin{pmatrix}
        \sqrt{T}\cos{\theta_{m}} & \sqrt{T}\sin{\theta_{m}} & 0 & 0 &\sqrt{1-T}\cos{\theta_{m}} & \sqrt{1-T}\sin{\theta_{m}} \\
        \sqrt{1-T}\cos{\theta_{lb}} & \sqrt{1-T}\sin{\theta_{lb}} & 0 & 0 & -\sqrt{T}\cos{\theta_{lb}} & -\sqrt{T}\sin{\theta_{lb}} \\
      0 &0 & \cos{\theta_{lc}} & \sin{\theta_{lc}} & 0 & 0 
      \end{pmatrix}.
\end{split}
\end{equation}
\end{widetext}
\end{appendix}
\balance

\end{document}